\documentclass[12pt]{iopart}      
\begin{document}
\title[]{Gauge anomaly with vector and axial-vector fields 
in six dimensional curved space}
\author{Satoshi Yajima$^*$, Kohei Eguchi, Makoto Fukuda, \\
and Tomonori Oka}
\address{Department of Physics, Kumamoto University, 
2-39-1 Kurokami, Chuo-ku, Kumamoto 860-8555, Japan\\
E-mail: yajima-s@kumamoto-u.ac.jp}
\begin{abstract}
Imposing the conservation equation of the vector current 
for a fermion of spin $\frac{1}{2}$ at the quantum level, 
a gauge anomaly for the fermion 
coupling with non-Abelian vector and axial-vector fields 
in six-dimensional curved space is expressed in tensorial form. 
The anomaly consists of terms that resemble the chiral U(1) anomaly
and the commutator terms that disappear if the axial-vector field is Abelian.
\end{abstract}
%
%

\section{Introduction}
The gauge anomaly breaks down the unitarity of the quantum field theory,  
and then one cannot calculate higher-order quantum corrections 
in a consistent manner.
The cancellation of the anomaly is a stringent condition 
on the fermion multiplets allowed in the consistent model. 
It is meaningful to clarify the form of the anomaly 
in general non-Abelian gauge theory.
  
The gauge anomaly arises from the breaking of a certain local gauge symmetry.
The concrete form of the anomaly is obtained by calculating some one-loop diagrams of fermions 
in four dimensions~\cite{Adler,BJ,Bardeen}, and is also derived from the chiral gauge transformation of 
the path integral measure for fermions interacting 
with boson fields~\cite{Fujikawa1,Fujikawa2,Fujikawa3,FS}.  
The formal expression of the anomaly 
in Gaussian cut-off regularization is described by the heat
kernel~\cite{Schwinger,DeWitt}.

We consider the action $\tilde{S}$ in the model in which
the fermion $\psi$ of spin $\frac{1}{2}$ interacts 
with the (polar-)vector field $\tilde{V}_{\mu}$ 
and the axial-vector field $\tilde{A}_{\mu}$ 
in even $2n$-dimensional curved space,
\begin{eqnarray}
&& \tilde{S} = \int d^{2n}x\, 
h\, \bar{\psi}\, {\rm i} \gamma^{\mu} (\nabla_{\mu} - {\rm i} \tilde{V}_{\mu} - {\rm i} \gamma_{2n+1} \tilde{A}_{\mu}) \, \psi , \nonumber \\
&& \nabla_{\mu} \psi = \partial_{\mu} \psi
+ \frac{1}{4}\, \omega^{kl}{}_{\mu}\, \gamma_{kl}\, \psi , \qquad
\tilde{V}_{\mu} = \tilde{V}^a_{\mu}\, T^a , \qquad \tilde{A}_{\mu} = \tilde{A}^a_{\mu}\, T^a, \nonumber \\
&& \gamma_{k_1 \cdots k_j} = \gamma_{[k_1} \cdots \gamma_{k_j]}, 
\qquad \gamma_{2n+1} = {\rm i}^n \gamma^1 \gamma^2 \cdots \gamma^{2n},
\label{eq:original_action} 
\end{eqnarray}
where $\omega^{kl}{}_{\mu}$ is the spin connection, and 
$h=\det{h^k{}_\mu}$, in which $h^k{}_{\mu}$ is a vielbein in curved
space. 
Note that the Euclidean metric tensor is 
$g_{\mu\nu} = h^k{}_{\mu} h^l{}_{\nu} \eta_{kl}$ 
with $\eta_{kl} = -\, \delta_{kl}$ in flat tangent space.
Moreover, $T^a$ denotes the Hermitian representation matrix 
of the Lie algebra of a non-Abelian gauge group. Both components
$\tilde{V}^a_{\mu}$ and $\tilde{A}^a_{\mu}$ of the boson fields are real.
The action $\tilde{S}$ is invariant under the infinitesimal local gauge transformation,
\begin{eqnarray}
\fl \qquad && \delta\psi(x) = ({\rm i} \alpha(x) + {\rm i} \beta(x) \gamma_{2n+1}) \psi(x), \qquad 
\delta \bar{\psi}(x) = \bar{\psi}(x) ( - {\rm i} \alpha(x) + {\rm i} \beta(x) \gamma_{2n+1}) , \nonumber \\
\fl \qquad && \delta\tilde{V}_{\mu}(x) =  \partial_{\mu} \alpha(x) + {\rm i} [ \alpha(x) , \tilde{V}_{\mu}(x)] + {\rm i} [ \beta(x) , \tilde{A}_{\mu}(x)] , \nonumber \\
\fl \qquad && \delta\tilde{A}_{\mu}(x) = \partial_{\mu} \beta(x) + {\rm i} [ \beta(x) , \tilde{V}_{\mu}(x)] + {\rm i} [ \alpha(x) , \tilde{A}_{\mu}(x)] ,
\label{eq:GT} 
\end{eqnarray}
where $\alpha(x) = \alpha^a(x) T^a$ and $\beta(x) = \beta^a(x) T^a$, in which $\alpha^a(x)$ and $ \beta^a(x)$ are real parameters.
The Dirac operator $\gamma^{\mu} (\nabla_{\mu} - {\rm i} \tilde{V}_{\mu} - {\rm i} \gamma_{2n+1} \tilde{A}_{\mu})$ in the action $\tilde{S}$ is not Hermitian.  
If one rotates the axial-vector $\tilde{A}_{\mu}$ to an imaginary field ${\rm i} A_{\mu}$ in which $A^a_{\mu}$ is real, then
the Dirac operator
${\ooalign{\hfil/\hfil\crcr${D}$}} \equiv \gamma^{\mu} (\nabla_{\mu} - {\rm i} \tilde{V}_{\mu} + \gamma_{2n+1} A_{\mu}) $ 
becomes Hermitian.  However, the rotation of $\tilde{A}_{\mu}$ spoils the axial-part of the gauge transformation, and 
the gauge symmetry for the axial-vector gauge field parametrized by $\beta \gamma_{2n+1}$ breaks in the path integral.
For simplicity, rewriting the vector field as $- {\rm i} \tilde{V}^a_{\mu} \equiv V^a_{\mu}$, which is purely imaginary,
the action is replaced by
\begin{eqnarray}
&& S = \int d^{2n}x\, 
h\, \bar{\psi}\, {\rm i} {\ooalign{\hfil/\hfil\crcr${D}$}}\, \psi ,
\qquad 
{\ooalign{\hfil/\hfil\crcr${D}$}}
= \gamma^{\mu} \nabla_{\mu} + Y \equiv \gamma^{\mu} D_{\mu}, \nonumber \\
&& Y = \gamma^{\mu} V_{\mu} + \gamma_{2n+1} \gamma^{\mu} A_{\mu}, \qquad
V_{\mu}= V^a_{\mu}\, T^a , \qquad A_{\mu} = A^a_{\mu}\, T^a .
\label{eq:action} 
\end{eqnarray}

In supergravity coupled with super Yang-Mills theory~\cite{SgSYM1,SgSYM2}, 
the Lagrangian contains four-fermion interactions,
which are regarded as some two-fermion interactions 
with bosonic background fields expressed by odd-order tensors. 
The completely antisymmetric part of the highest order tensor
should be rewritten as an axial-vector 
by contracting its tensor with the Levi-Civita symbol.
The vector and the axial-vector parts in the two-fermion interactions 
can be absorbed in the vector and the axial-vector gauge fields.
(Other order tensors may be treated in future work.)
The concrete form of the gauge anomaly in the model may be directly calculated by using the heat kernel.

It is shown in Sect.~2 that the gauge anomaly stems from the Jacobian for the functional measure 
of the fermion in the path integral under chiral transformation.
The heat kernel is introduced in Sect.~3, 
in order to give the explicit form of the anomaly 
in four and six dimensions in Sects.~4 and 5, respectively. 
Section 6 is devoted to the discussion.

\section{Gauge anomaly}

The path integral, in which $V_{\mu}$ and $A_{\mu}$ 
are regarded as background fields, is given by
\begin{equation}
W(V_{\mu},A_{\mu}) =
\ln \int {\cal D} \bar{\psi} {\cal D} \psi
\exp S(V_{\mu},A_{\mu}, \bar{\psi} , \psi).
\end{equation}
The Ward--Takahashi identity in the path integral
can be expressed in the following form, 
due to the replacement of the fermion corresponding to 
the infinitesimal transformation of the fermion in $(\ref{eq:GT})$;
$\psi^{\prime} = (1 + {\rm i} \alpha + {\rm i} \beta \gamma_{2n+1}) \psi$,
$\bar{\psi}^{\prime} = \bar{\psi} (1 - {\rm i} \alpha + {\rm i} \beta \gamma_{2n+1})$,
\begin{eqnarray}
\fl \qquad 0 &=& \ln \int {\cal D} \bar{\psi}^{\prime} {\cal D} \psi^{\prime}
\exp S(V_{\mu},A_{\mu}, \bar{\psi}^{\prime} , \psi^{\prime})
- \ln \int {\cal D} \bar{\psi} {\cal D} \psi
\exp S(V_{\mu},A_{\mu}, \bar{\psi} , \psi) \nonumber \\
\fl \qquad &=& \ln \int {\cal D} \bar{\psi} {\cal D} \psi {\Big[
\ln{J_{\bar{\psi}}} + \ln{J_{\psi}} - \int d^{2n}x\, 
h\, \bar{\psi}\, \gamma^{\mu} D_{\mu} (\alpha(x) + \beta(x) \gamma_{2n+1})\,
\psi \Big]} e^S ,
\label{eq:WT}
\end{eqnarray}
where $J_{\psi}$ and $J_{\bar{\psi}}$ are the Jacobians for the transformation of ${\cal D} \psi$ and ${\cal D} \bar{\psi}$.
To analyze the Jacobians,
we use the complete set of eigenfunctions $\{ \varphi_{n} \}$ of the Hermitian {\ooalign{\hfil/\hfil\crcr$D$}}:
\begin{equation}
{\ooalign{\hfil/\hfil\crcr$D$}} \varphi_n(x)
=
\lambda_n \varphi_n(x),
\qquad
\int d^{2n}x\ h(x) \varphi_m^\dagger(x) \varphi_n(x)
= \delta_{mn}.
\label{eq:eigen}
\end{equation}
Expanding the fields $\bar{\psi}$ and $\psi$ by $\{ \varphi_{n} \}$ as
\begin{equation}
 \psi(x) = \sum_n a_n\, \varphi_n(x), \qquad 
\bar{\psi} = \sum_n \bar{b}_n\, \varphi^{\dag}_n(x),
\label{eq:expansion}
\end{equation}
with Grassmann number coefficients $a_n$ and $\bar{b}_n$, we define the path integral measures by 
\begin{equation}
{\cal D} \bar{\psi} {\cal D} \psi = \prod_n d\bar{b}_n\,  da_n.
\label{eq:expansion}
\end{equation}
Then the Jacobian factors $\ln{J_{\psi}}+\ln{J_{\bar{\psi}}}$ in $(\ref{eq:WT})$ are expressed as  
\begin{equation}
\ln{J_{\psi}}+\ln{J_{\bar{\psi}}}
= -2{\rm i} \sum_n
\int d^{2n}x\ \beta^a\,  \varphi_n^\dagger(x) T^a \gamma_{2n+1} \varphi_n(x).
\label{eq:J-cov}
\end{equation}
The relation $(\ref{eq:WT})$ is rewritten 
by separating terms containing the real parameters $\alpha^a$ and $\beta^a$:
\begin{eqnarray}
&& D_{\mu} \langle \bar{\psi} T^a \gamma^{\mu} \psi \rangle (x) = 0,\qquad 
D_{\mu} \langle \bar{\psi} T^a \gamma_{2n+1} \gamma^{\mu} \psi \rangle (x) = G^{(2n)a}(x),  \nonumber \\
&& 
G^{(2n)a}(x) =
 -2{\rm i} \sum_n \varphi_n^\dagger(x) T^a \gamma_{2n+1} \varphi_n(x).
\label{eq::naiveGA}
\end{eqnarray}
Note that the gauge anomaly $G^{(2n)a}$ does not break the conservation law of the vector current, 
which is related to Noether's theorem with respect to the gauge transformation parametrized by $\alpha(x)$. 

\section{Heat kernel}
Since the expression $(\ref{eq::naiveGA})$ of the anomaly is ill-defined,
it should be regularized in order
to calculate concretely in the tensorial form.
For this purpose, the Gaussian cut-off regularization is adopted~\cite{TE}:
\begin{eqnarray}
\fl \qquad G^{(2n)a}(x)
&=& -2{\rm i} \lim_{t \to 0}\sum_n \varphi_n^\dagger(x) T^a \gamma_{2n+1}\,
{\rm e}^{-t{\ooalign{\hfil/\hfil\crcr${D}$}}^2} \varphi_n(x) \nonumber \\
\fl \qquad &=& -2{\rm i} \lim_{t \to 0} \lim_{x^{\prime} \to x}
{\rm Tr} \left( T^a \gamma_{2n+1} \, 
{\rm e}^{-t{\ooalign{\hfil/\hfil\crcr${D}$}}^2} 
|h(x)|^{- \frac{1}{2}}\, |h(x')|^{- \frac{1}{2}}\, 
\delta^{(2n)}(x,x^{\prime}) \right) \nonumber \\
\fl \qquad &\equiv& -2{\rm i} \lim_{t \to 0} \lim_{x^{\prime} \to x}
{\rm Tr} \left( T^a \gamma_{2n+1} \, K^{(2n)}(x,x^{\prime};t) \right),
\label{eq:EA_GC}
\end{eqnarray}
where ${\rm Tr}$ denotes the trace over 
both spinor indices of $\gamma$-matrices and internal indices of $T^a$. 
The heat kernel $K^{(2n)}(x,x^{\prime};t)$ introduced in $(\ref{eq:EA_GC})$ satisfies  
the heat equation and the boundary condition with respect to $t$~\cite{Schwinger,DeWitt}:
\begin{eqnarray}
&& {\partial \over \partial t}\, K^{(2n)}(x,x';t) =
-\,  {\ooalign{\hfil/\hfil\crcr${D}$}}^2 K^{(2n)}(x,x';t), \label{eq:K-eq} \\
&& K^{(2n)}(x,x';0) =
{\bf 1}\, |h(x)|^{- \frac{1}{2}}\, 
|h(x')|^{- \frac{1}{2}}\, \delta^{(2n)}(x,x'), \label{eq:K-con} 
\end{eqnarray}
where ${\bf 1}$ is a unit matrix not only acting on the spinor but also acting on $T^a$, 
and $\delta^{(d)}(x,x')$ means the invariant $\delta$ function.
Moreover, the square of ${\ooalign{\hfil/\hfil\crcr${D}$}}$ for $\psi$ is rewritten as
\begin{eqnarray}
\fl \qquad && {\ooalign{\hfil/\hfil\crcr${D}$}}^2 
= D_{\mu} D^{\mu} + 2 Q^{\mu} D_{\mu} + Z
= \tilde{D}_{\mu} \tilde{D}^{\mu} + X, \nonumber \\
\fl \qquad && Q_{\mu} = \frac{1}{2} \{ \gamma_{\mu} , Y \},  \qquad 
Z = \frac{1}{2}\, \gamma^{\mu\nu}\, [ \nabla_{\mu}, \nabla_{\nu} ] 
+ \gamma^{\mu} \nabla_{\mu} Y + Y^2,  \nonumber \\
\fl \qquad && \tilde{D}_{\mu} = \nabla_{\mu} + Q_{\mu}, \qquad
X = Z - \nabla_{\mu} Q^{\mu} - Q_{\mu} Q^{\mu}, \qquad
[ \tilde{D}_{\mu}, \tilde{D}_{\nu} ]\psi = \Lambda_{\mu\nu}\, \psi.
\label{eq:QXA}
\end{eqnarray}
The matrix-valued quantities $Q_{\mu}$, $X$, and $\Lambda_{\mu\nu}$ 
are expressed from $(\ref{eq:QXA})$
in the following form:
\begin{eqnarray}
\fl \qquad &&Q_{\mu} = V_{\mu} - \gamma_{2n+1}\, \gamma_{\mu\rho} A^{\rho}, \qquad
F_{\mu\nu} = \partial_{\mu} V_{\nu} - \partial_{\nu} V_{\mu}
+ [V_{\mu} , V_{\nu} ], \nonumber \\
\fl \qquad &&X = -\, {1 \over 4}\, R + 2(n-1) \, A_{\mu} A^{\mu} 
- \gamma_{2n+1}\, A^{\mu}{}_{;\mu}
+ \gamma^{\mu\nu} \left( {1 \over 2}\, F_{\mu\nu} 
+ {2n-3 \over 2}\, [A_{\mu} , A_{\nu} ] \right), \nonumber \\
\fl \qquad &&\Lambda_{\mu\nu} 
= {1 \over 4}\, \gamma^{\rho\sigma} R_{\rho\sigma\mu\nu}
+ F_{\mu\nu} - [A_{\mu} , A_{\nu} ] 
- 2\, \gamma_{\mu\nu} A_{\rho} A^{\rho} 
+ 2\, \gamma_{[\mu|}{}^{\rho} \{ A_{|\nu]} , A_{\rho} \} \nonumber \\
\fl \qquad &&\qquad\quad +\, 2\, \gamma_{2n+1}\, \gamma_{[\mu|\rho}  
A^{\rho}{}_{;|\nu]}
- 2 \, \gamma_{\mu\nu\rho\sigma} A^{\rho} A^{\sigma}, 
\label{Lam}
\end{eqnarray}
where $R_{\alpha\beta\mu\nu}$ is the Riemann curvature tensor,
and the semi-colon `$;$' means
the covariant differentiation preserving the vector gauge and the gravitational symmetries
(e.g. $A^{\rho}{}_{;\nu} = \nabla_{\nu} A^{\rho} + [ V_{\nu} , A^{\rho} ]$).
The totally antisymmetric product
$\gamma_{\mu\nu\rho\sigma}$ in the last term of $\Lambda_{\mu\nu}$ 
is expressed as
$- \epsilon_{\mu\nu\rho\sigma} \gamma_5$ 
and $- \frac{\rm i}{2} \epsilon_{\mu\nu\rho\sigma\kappa\lambda} 
\gamma_7 \gamma^{\kappa\lambda}$ 
in four and six dimensions, respectively.

It is difficult to solve the equation $(\ref{eq:K-eq})$ 
of the heat kernel 
for the fermion interacting 
with general background fields. 
In order to perform the concrete calculation, the following ansatz by DeWitt is 
applied to the heat kernel~\cite{DeWitt}, 
which satisfies the condition $(\ref{eq:K-con})$:
\begin{equation}
 K^{(2n)}(x,x';t) \sim {\Delta^{1/2}(x,x') \over (4 \pi t)^n}\, 
{\rm exp} \left({\sigma(x,x') \over 2t} \right) 
\sum_{q=0}^{\infty} a_q(x,x')\, t^q,  
\label{eq:DWansatz}
\end{equation}
where $\sigma(x,x')$ is half of the square of the geodesic distance 
between $x$ and $x'$, 
$\Delta(x,x') = |h(x)|^{-1}\, 
|h(x')|^{- 1} \det \sigma_{;\mu\nu^{\prime}}$, 
and $a_q(x,x')$ stand for bispinors. 
The coincidence limit of $a_n$ appears in the formal expression of the
anomaly, and is defined by
$\lim_{x'\to x}a_n(x,x') \equiv [a_n](x)$. In particular, 
 $[\sigma]=[\sigma_{;\mu}] =0, 
[\sigma_{;\mu\nu}]= g_{\mu\nu}, [\Delta] =1$, and $[a_0] ={\bf 1}$. 

\section{Gauge anomaly in four dimensions}
Substituting the ansatz $(\ref{eq:DWansatz})$ into $(\ref{eq:EA_GC})$,
the gauge anomaly in $2n$ dimensions 
is derived from $[a_n]$,~\cite{Kimura}
\begin{equation}
G^{(2n)a}(x)
= {-2{\rm i} \over (4 \pi)^n} {\rm Tr} 
\Big\{T^a \, \gamma_{2n+1}\, [a_n] \Big\}(x). \label{eq:GA}  
\end{equation}
If only $T^a$ written visibly in (\ref{eq:GA}) is a unit matrix,\footnote{
The matrix $T^a$ noted here does not mean that contained in $[a_n]$.}  
then 
$G^{(2n)a}$ becomes the chiral U(1) anomaly.  
The concrete form of $[a_2]$ and $[a_3]$ is given as follows~\cite{Gilkey,Avramidi,Ven}:\footnote{
In $[a_3]$, some terms should be described by a commutator and some anticommutators, 
because $A_{\mu}$ and $F_{\mu\nu}$ in $X$, $\Lambda_{\mu\nu}$ and their
derivatives do not commute with $T^a$ in (\ref{eq:GA}). If $X$ and $\Lambda_{\mu\nu}$ commute
with $T^a$, then a commutator disappears, 
and anticommutators of two quantities double the product of them, in the trace formula.}  
\begin{eqnarray}
\fl \quad && [a_2] = {1 \over 12} \Lambda_{\mu\nu} \Lambda^{\mu\nu}
+ {1 \over 6} X_{!\mu}{}^{\mu}
+ {1 \over 2} \left({1 \over 6} R + X \right)^2 
+ \cdots , \label{eq:a2} \\
\fl \quad && [a_3] = {1 \over 60} \Big( 
-{1 \over 3} \left(X_{!\mu}{}^{\mu}{}_{\nu}{}^{\nu}
+ X_{!\mu\nu}{}^{\nu\mu} + X_{!\mu\nu}{}^{\mu\nu}\right) 
- {1 \over 3} \Lambda_{\mu\nu}{}^{!\nu}
\Lambda^{\mu\rho}{}_{!\rho}
- {4 \over 3} \Lambda_{\mu\nu!\rho} 
\Lambda^{\mu\nu!\rho} \nonumber \\
\fl \quad && \phantom{[a_3] ={1 \over 60} \Big( }
- 2 \{ \Lambda^{\mu\nu} ,
\Lambda_{\mu\rho}{}^{!\rho}{}_{\nu} \} 
- {10 \over 3} R^{\mu\nu} 
\Lambda_{\mu\rho} \Lambda_{\nu}{}^{\rho}
+ R^{\mu\nu\rho\sigma} \Lambda_{\mu\nu}
\Lambda_{\rho\sigma}
- 6 \Lambda_{\mu}{}^{\nu}
\Lambda_{\nu}{}^{\rho} \Lambda_{\rho}{}^{\mu} \Big)
 \nonumber \\
\fl \quad && \phantom{[a_3] =}+ {1 \over 12} \Big\{ {1 \over 6} R + X ,
-\, {1 \over 2} \Lambda_{\mu\nu} \Lambda^{\mu\nu}
- X_{!\mu}{}^{\mu} - {1 \over 5} R_{;\mu}{}^{\mu} 
+ {1 \over 30} R_{\mu\nu} R^{\mu\nu} 
- {1 \over 30} R_{\mu\nu\rho\sigma} R^{\mu\nu\rho\sigma} \Big\}
\nonumber \\
\fl \quad && \phantom{[a_3] =}- {1 \over 36} [ X_{!\mu} , \Lambda^{\mu\nu}{}_{!\nu} ]
- {1 \over 12} \left({1 \over 6} R + X \right)_{!\rho}
\left({1 \over 6} R + X \right)^{!\rho}
- {1 \over 6} \left({1 \over 6} R + X \right)^3 + \cdots,
\label{eq:a3}
\end{eqnarray}
where the exclamation mark `$!$' means  
a new covariant differentiation $\tilde{D}_{\mu}$ introduced in (\ref{eq:QXA}), 
and terms without matrices are omitted from (\ref{eq:a2})
and (\ref{eq:a3}). 
The tensorial form of the gauge anomaly in four dimensions is written as
\begin{eqnarray}
\fl \quad G^{(4)a}
&=& {-2{\rm i} \over (4 \pi)^2} {\rm tr}\, T^a \bigg[  
\epsilon^{\alpha\beta\gamma\delta} \Big\{- {1 \over 48} 
R_{\alpha\beta\mu\nu} R_{\gamma\delta}{}^{\mu\nu}
- {1 \over 2} V_{\alpha\beta} V_{\gamma\delta}
- {1 \over 6} A_{\alpha\beta} A_{\gamma\delta} \nonumber \\
\fl \quad &&\qquad \qquad  
+\, {4 \over 3}(V_{\alpha\beta}  A_{\gamma}  A_{\delta}
+ A_{\alpha} V_{\beta\gamma} A_{\delta}
+ A_{\alpha} A_{\beta} V_{\gamma\delta})
- {16 \over 3} A_{\alpha} A_{\beta} A_{\gamma} A_{\delta} \Big\} 
\nonumber \\
\fl \quad &&\qquad \qquad  
-\, {2 \over 3} A^{\mu}{}_{;\mu\nu}{}^{\nu}
+ {1 \over 3} R A^{\mu}{}_{;\mu}
- {2 \over 3} R^{\mu\nu} A_{\mu;\nu} 
- {4 \over 3} [A_{\mu} , V^{\mu\nu}{}_{;\nu}] 
+ {1 \over 3} [A_{\mu\nu} , V^{\mu\nu}] \nonumber \\
\fl \quad &&\qquad \qquad 
-\, {4 \over 3} \{ \{ A_{\mu} , A_{\nu} \} , A^{\mu;\nu} \}
+ {2 \over 3} \{ A_{\mu} A^{\mu} ,A^{\nu}{}_{;\nu} \}     
- 4 A_{\nu} A^{\mu}{}_{;\mu} A^{\nu}
 \bigg] \label{eq:famous} \\
\fl \quad &=&{-2{\rm i} \over (4 \pi)^2} {\rm tr}\, T^a \bigg[  
- {1 \over 48} \epsilon^{\alpha\beta\gamma\delta}
R_{\alpha\beta\mu\nu} R_{\gamma\delta}{}^{\mu\nu}
- {1 \over 2} \epsilon^{\alpha\beta\gamma\delta}
F_{\alpha\beta} F_{\gamma\delta} 
+ \Big( - {2 \over 3} A^{\delta;\mu}{}_{\mu}  \nonumber \\
\fl \quad &&\qquad \qquad 
+\, {1 \over 3} R A^{\delta} 
- 2 [ A_{\mu} , F^{\mu\delta} ]
- {1 \over 3} \epsilon^{\alpha\beta\gamma\delta}
\{ A_{\alpha;\beta} , A_{\gamma} \}
- {4 \over 3} \{ A_{\mu} A^{\mu} , A^{\delta} \}
\Big)_{;\delta} \nonumber \\
\fl \quad &&\qquad \qquad 
+\, {1 \over 2} \epsilon^{\alpha\beta\gamma\delta}
[\{F_{\alpha\beta} , A_{\gamma} \} , A_{\delta} ]
+ {2 \over 3} [A_{\mu},[A^{\mu} , A^{\nu}{}_{;\nu}]]
- {2 \over 3} [ A_{\mu;\nu} , [A^{\mu} , A^{\nu}]]
\nonumber \\
\fl \quad &&\qquad \qquad  
+\, {1 \over 3} \epsilon^{\alpha\beta\gamma\delta}
[ A_{\alpha} , A_{\beta} A_{\gamma} A_{\delta} ] 
 \bigg] , \label{eq:GA4}
\end{eqnarray}
where 
${\rm tr}$ runs only over the internal indices of $T^a$, and
\begin{eqnarray*}
&&V_{\mu\nu} \equiv \partial_{\mu} V_{\nu} - \partial_{\nu} V_{\mu}
+ [V_{\mu},V_{\nu}] + [A_{\mu},A_{\nu}] 
= F_{\mu\nu} + [A_{\mu},A_{\nu}]\,, \\
&&A_{\mu\nu} \equiv \partial_{\mu} A_{\nu} - \partial_{\nu} A_{\mu}
+ [V_{\mu} , A_{\nu}] - [V_{\nu} , A_{\mu}] 
= A_{\nu;\mu} - A_{\mu;\nu}\,. \label{CU1A6}
\end{eqnarray*}

The expression (\ref{eq:famous}) of the gauge anomaly 
is well known~\cite{Bardeen,GA1,GA2,GA3a,GA3b,GA4}.
The leading terms without $A_\mu$ in the anomaly are represented by 
$\epsilon^{\alpha\beta\gamma\delta} R_{\alpha\beta\mu\nu}
R_{\gamma\delta}{}^{\mu\nu}$ 
and $\epsilon^{\alpha\beta\gamma\delta} F_{\alpha\beta}
F_{\gamma\delta}$, 
which are shown in the chiral U(1) anomaly.
The terms containing $A_\mu$ in (\ref{eq:GA4}) consist of the total derivative terms
and the commutator terms of $F_{\alpha\beta}$, $A_{\mu}$, and their derivatives.

\section{Gauge anomaly in six dimensions}
Since the chiral U(1) anomaly in the vector and axial-vector model in six dimensions 
has been derived \cite{CA6}, 
we can present the leading and the total derivative terms of the gauge anomaly:\footnote{
In expression (11) in Ref.~\cite{CA6}, the coefficients of 
$+ {1 \over 40} R^{\mu\nu\kappa\lambda} R_{\mu\nu\kappa\lambda} A^{\rho}$, 
$- {1 \over 2} \{A_{\mu:\nu}, A^{\nu:\rho} \} A^{\mu}$, and 
$- {49 \over 30} F^{\nu\rho} [A_{\mu} A^{\mu} , A_{\nu} ]$ on the 6th, 12th, and 15th lines 
should be replaced with $- {7 \over 1440} R^{\mu\nu\kappa\lambda} R_{\mu\nu\kappa\lambda}
A^{\rho}$, $- {17 \over 30} \{A_{\mu:\nu}, A^{\nu:\rho} \} A^{\mu}$, 
and $- {3 \over 10} F^{\nu\rho} [A_{\mu} A^{\mu} , A_{\nu} ]$, respectively,
and the term $-\, {1 \over 15} \{A^{\mu:\rho}, A_{\mu:\nu} \} A^{\nu}$ 
on the 12th line should be eliminated. 
The modified chiral U(1) anomaly is given from $(\ref{CU1A6})$ in this article.}
\begin{eqnarray}
\fl G^{(6)a} &=& 
{- 2{\rm i} \over 8 \pi^3}\, {\rm tr}\,T^a \bigg[ - {{\rm i} \over 8}\,
\epsilon^{\alpha\beta\gamma\delta\kappa\lambda}
\left({1 \over 48} R_{\alpha\beta\rho\sigma} R_{\gamma\delta}{}^{\rho\sigma}
+ {1 \over 6} F_{\alpha\beta} F_{\gamma\delta} \right) F_{\kappa\lambda} 
\nonumber \\
\fl &&+\, \left({1 \over 180} \left(A^{\mu}{}_{;\mu\nu}{}^{\nu\rho} 
+ A^{\mu}{}_{;\mu\nu}{}^{\rho\nu} 
+ A^{\mu}{}_{;\mu}{}^{\rho\nu}{}_{\nu} \right)
- {1 \over 72} R A_{\mu}{}^{;\mu\rho} 
- {1 \over 90} R^{\rho\mu} A_{\mu;\nu}{}^{\nu}
\right. \nonumber \\
\fl &&\quad +\, {2 \over 45} R_{\mu\nu} A^{\mu;\nu\rho} 
- {1 \over 36} R^{\mu\nu\lambda\rho} A_{\mu;\lambda\nu}  
+\, {1 \over 120} \left(-\, R^{;\rho} A_{\mu}{}^{;\mu}
+ R_{;\mu} A^{\rho;\mu} + R_{;\mu} A^{\mu;\rho} \right) 
\nonumber \\
\fl &&\quad 
+ {1 \over 30} \left(-\, R^{\rho\mu;\nu} + R^{\mu\nu;\rho} \right)
A_{\nu;\mu}
+ {1 \over 288} R^2 A^{\rho}
- {1 \over 72} R R^{\mu\rho} A_{\mu}
+\, {1 \over 90} R_{\mu\nu} R^{\mu\rho} A^{\nu} 
\nonumber \\ 
\fl &&\quad +\, {1 \over 36} R_{\mu\nu} R^{\mu\lambda\nu\rho} A_{\lambda} 
- {1 \over 180} R^{\mu\nu\kappa\rho} R_{\mu\nu\kappa\lambda} 
A^{\lambda}
- {7 \over 1440} R^{\mu\nu\kappa\lambda} R_{\mu\nu\kappa\lambda} 
A^{\rho} \nonumber \\
\fl &&\quad
-\, {1 \over 24} \{ F^{\mu\nu}\,,\, \{ F_{\mu\nu}\,,\, A^{\rho} \} \}
+\, {1 \over 12} \{ F^{\mu\nu}\,,\, \{ F^{\rho}{}_{\nu}\,,\, A_{\mu} \} \}
- {{\rm i} \over 24} \, \epsilon^{\alpha\beta\gamma\delta\kappa\rho}
\{F_{\alpha\beta}\,,\, A_{\gamma;\delta} A_{\kappa} \} 
\nonumber \\
\fl &&\quad +\, {11 \over 30} \{ A_{\mu} A^{\mu} , A^{\rho;\nu}{}_{\nu} \}
+ {2 \over 3} \{ A_{\mu} A^{\mu} , A_{\nu}{}^{;\nu\rho} \}
- {19 \over 30} \{ A_{\mu} A^{\mu} , A_{\nu}{}^{;\rho\nu} \} 
 \nonumber \\
\fl \quad &&\quad 
-\, {1 \over 20} \{ \{ A^{\rho}, A^{\mu} \} , A^{\nu}{}_{;\mu\nu} \} 
+ {3 \over 20} \{ \{ A^{\rho} , A^{\mu} \} , A_{\mu;\nu}{}^{\nu} \}
+ {1 \over 60} \{ \{ A^{\rho} , A^{\mu} \} , A^{\nu}{}_{;\nu\mu} \} 
\nonumber \\
\fl &&\quad 
+\, {1 \over 6} \{ \{A^{\mu}, A^{\nu} \} , A_{\mu;\nu}{}^{\rho} \}
+ {1 \over 30} \{ \{A^{\mu}, A^{\nu} \} , A^{\rho}{}_{;\mu\nu} \}
- {11 \over 60} \{ \{A^{\mu}, A^{\nu}\} , A_{\mu}{}^{;\rho}{}_{\nu} \} 
 \nonumber \\
\fl &&\quad 
- {17 \over 60} \{ \{A_{\mu;\nu} , A^{\nu;\rho} \} , A^{\mu} \}
- {1 \over 60} \{ \{A^{\rho;\mu} , A_{\mu:\nu} \} , A^{\nu} \}
+ {2 \over 5} \{ \{A^{\rho;\nu} , A_{\mu;\nu} \} , A^{\mu} \} 
 \nonumber \\
\fl &&\quad - {1 \over 20} \{ \{ A^{\nu;\rho} , A^{\mu}{}_{;\mu} \} , A_{\nu} \}
+\, {1 \over 30} \{ \{ A^{\rho;\nu} , A^{\mu}{}_{;\mu} \} , A_{\nu} \} 
 \nonumber \\
\fl &&\quad 
- {1 \over 30} \{ A^{\mu}{}_{;\mu} A^{\nu}{}_{;\nu} , A^{\rho} \}
+ {1 \over 5} \{ A^{\mu;\nu} A_{\mu;\nu} , A^{\rho} \}
- {1 \over 15} \{ A^{\mu;\nu} A_{\nu;\mu} , A^{\rho} \} \nonumber \\
\fl &&\quad 
-\, {3 \over 20} \{F^{\nu\rho}, [A_{\mu} A^{\mu} , A_{\nu} ] \}
- {1 \over 5} \{ F_{\mu\nu}, \{A^{\mu} A^{\nu} , A^{\rho} \} \} 
- {1 \over 15} \{ F_{\mu\nu}, A^{\mu} A^{\rho} A^{\nu} \} \nonumber \\
\fl &&\quad   
-\, {1 \over 18} R A_{\mu} A^{\mu} A^{\rho}
+\, {32 \over 45} R^{\rho\mu} A_{\mu} A_{\nu} A^{\nu} 
- {1 \over 15} R^{\rho\nu} A_{\mu} A_{\nu} A^{\mu}
- {1 \over 45} R^{\mu\nu} A_{\mu} A_{\nu} A^{\rho} \nonumber \\
\fl &&\quad 
-\, {{\rm i} \over 20}\, \epsilon^{\alpha\beta\gamma\delta\kappa\rho}
A_{\alpha} A_{\beta} A_{\gamma} A_{\delta;\kappa}
+\,  {1 \over 5} A_{\mu} A^{\mu} A_{\nu} A^{\nu} A^{\rho} \qquad \qquad \qquad \qquad \qquad \qquad \qquad 
\nonumber \\
\fl &&\quad \left.
- {1 \over 3} A_{\mu} A_{\nu} A^{\mu} A^{\nu} A^{\rho} 
+ {2 \over 5} A_{\mu} A_{\nu} A^{\nu} A^{\mu} A^{\rho} \right)_{;\rho} 
 \bigg] 
+ G_{\rm com}^{a} , \label{eq:A6} 
\end{eqnarray}
where terms $G_{\rm com}^{a}$ containing the commutator 
are given in the following form: 
\begin{eqnarray}
\fl 
G_{\rm com}^{a} &=& 
{- {\rm i} \over 8 \pi^3}\, {\rm tr}\,T^a \bigg[
{1 \over 12} [[ F^{\rho\nu} , F_{\mu\nu}] , A^{\mu}]_{;\rho} 
-{1 \over 36} [ F_{\rho\mu}{}^{;\mu} , A_{\nu}{}^{;\nu\rho} +A^{\rho;\nu}{}_{\nu} + [ F^{\rho\nu} , A_{\nu} ] ] 
\nonumber \\
\fl &&
+ {1 \over 36} [ F^{\alpha\beta;\rho} , A_{\alpha;\rho\beta} + [ F_{\rho\alpha} , A_{\beta} ] ]
- {1 \over 36} R^{\alpha\beta} [ F_{\rho\alpha;\beta} , A^{\rho} ]
- {1 \over 72} R^{\alpha\beta\gamma\delta} [ F_{\alpha\beta;\gamma} , A_{\delta} ]
\nonumber \\
\fl &&+\,{\rm i}\, \epsilon^{\alpha\beta\gamma\delta\kappa\lambda} 
\Big( {1 \over 96} R_{\alpha\beta\rho\sigma} R_{\gamma\delta}{}^{\rho\sigma}
[ A_{\kappa} , A_{\lambda} ] 
+ {1 \over 48} [ F_{\alpha\beta} ,
F_{\gamma\delta} A_{\kappa} A_{\lambda}] \nonumber \\
\fl &&\qquad \qquad 
-\, {1 \over 144} [ F_{\alpha\beta} A_{\kappa} F_{\gamma\delta} , A_{\lambda} ]
- {1 \over 72} [ F_{\alpha\beta} A_{\kappa} , A_{\lambda} F_{\gamma\delta} ]
 \Big) \nonumber \\
\fl && 
-\, {3 \over 20} [ A^{\mu} , [ A_{\mu} ,
A_{\rho}{}^{;\rho\nu}{}_{\nu} ] ]
- {7 \over 45} [ A^{\mu} , [ A_{\mu} ,
A_{\rho;\nu}{}^{\nu\rho} ] ]
+ {1 \over 30} [ A^{\alpha} , [ A^{\beta} ,
A^{\rho}{}_{;\rho\beta\alpha} ] ]
 \nonumber \\
\fl &&
-\, {7 \over 90} [ A^{\alpha} , [ A^{\beta} ,
A_{\alpha;\beta\nu}{}^{\nu} ] ]
+ {11 \over 90} [ A^{\alpha} , [ A^{\beta} ,
A_{\alpha;\rho}{}^{\rho}{}_{\beta} ] ]
- {11 \over 270} [ A^{\alpha} , [ A^{\beta} ,
A^{\rho}{}_{;\alpha\beta\rho} ] ]
 \nonumber \\
\fl &&
+\, {29 \over 270} [ A^{\alpha} , [ A^{\beta} ,
A^{\rho}{}_{;\rho\alpha\beta} ] ]
+ {7 \over 180} [ [ A^{\alpha} , A^{\beta} ] ,
A_{\alpha;\beta\rho}{}^{\rho} ]
-\, {11 \over 180} [ [A^{\alpha} , A^{\beta} ] ,
A_{\alpha;\rho}{}^{\rho}{}_{\beta} ]
\nonumber \\
\fl &&
+ {11 \over 540} [ [ A^{\alpha} , A^{\beta} ] ,
A^{\rho}{}_{;\alpha\beta\rho} ]
- {29 \over 540} [ [ A^{\alpha}, A^{\beta} ] ,
A^{\rho}{}_{;\rho\alpha\beta} ]
 \nonumber \\
\fl &&+ {1 \over 60} [ A^{\alpha;\beta} , [ A_{\beta;\alpha} , 
A^{\rho}{}_{;\rho} ] ]
+ {41 \over 180} [ A^{\alpha;\beta} , [ A^{\rho}{}_{;\rho} ,
A_{\alpha;\beta} ] ]
+\, {13 \over 135} [ A_{\rho} , [ A^{\mu;\rho} , 
A^{\nu}{}_{;\nu\mu} ] ]
\nonumber \\
\fl &&
- {2 \over 45} [ A_{\rho} , [ A^{\rho;\mu} , 
A_{\mu;\nu}{}^{\nu} ] ] 
- {13 \over 18} [ A_{\rho} , [ A^{\rho;\mu} , 
A^{\nu}{}_{;\nu\mu} ] ]
-\, {4 \over 45} [ A^{\rho} , [ A_{\rho;\mu} , 
A_{\nu;}{}^{\mu\nu} ] ] \nonumber \\
\fl &&
- {5 \over 108} [ A^{\rho} , [ A^{\mu}{}_{;\mu} , 
A^{\nu}{}_{;\rho\nu} ] ] 
- {1 \over 540} [ A^{\rho} , [ A^{\mu}{}_{;\mu} , 
A^{\nu}{}_{;\nu\rho} ] ]
- {5 \over 36} [ A^{\rho} , [ A^{\mu}{}_{;\mu} , 
A_{\rho;\nu}{}^{\nu} ] ] \nonumber \\ 
\fl &&
+\, {1 \over 135} [ A^{\rho} , [ A^{\mu;\nu} , 
A_{\nu;\rho\mu} ] ] 
+ {1 \over 9} [ A^{\rho} , [ A^{\mu;\nu} , 
A_{\rho;\nu\mu} ] ]
+ {2 \over 45} [ A^{\rho} , [ A^{\mu;\nu} , 
A_{\mu;[\rho\nu]} ] ] \nonumber \\
\fl &&
+\, {1 \over 108} [ [ A^{\rho} , A^{\mu}{}_{;\mu} ] , 
A^{\nu}{}_{;\rho\nu} ]
+ {11 \over 108} [ [ A^{\rho} , A^{\mu}{}_{;\mu} ] , 
A^{\nu}{}_{;\nu\rho} ]
+\, {7 \over 36} [ [ A^{\rho} , A^{\mu}{}_{;\mu} ] , 
A_{\rho;\nu}{}^{\nu} ] \nonumber \\
\fl &&
-\, {1 \over 36} [ [ A_{\rho} , A^{\rho;\mu} ] , 
A_{\mu;\nu}{}^{\nu} ]
+ {13 \over 45} [ [ A_{\rho} , A^{\rho;\mu} ] , 
A^{\nu}{}_{;\nu\mu} ] 
+ {2 \over 45} [ [ A^{\rho} , A_{\rho;\mu} ] , 
A_{\nu;}{}^{\mu\nu} ] 
\nonumber \\
\fl &&
- {7 \over 108} [ [ A_{\rho} , A^{\mu;\rho} ] , 
A^{\nu}{}_{;\nu\mu} ] 
+\, {1 \over 60} [ [ A_{\rho} , A^{\mu;\rho} ] , 
A_{\mu;\nu}{}^{\nu} ]
- {11 \over 540} [ [ A^{\rho} , A^{\mu;\nu} ] , 
A_{\nu;\rho\mu} ]
\nonumber \\
\fl  &&
- {1 \over 45} [ [ A^{\rho} ,  A^{\mu;\nu} ] , 
A_{\rho;\nu\mu} ] 
+ {1 \over 180} [ [ A^{\rho} , A^{\mu;\nu} ] , 
A_{\mu;\rho\nu} ] 
- {1 \over 45} [ [ A^{\rho} ,  A^{\mu;\nu} ] , 
A_{\mu;\nu\rho} ] 
\nonumber \\  
\fl &&+ \bigg( {19 \over 90} [ A^{\alpha} , [ A_{\alpha} ,
A_{\beta}{}^{;\rho\beta} ] ]
- {7 \over 90} [ A^{(\rho} , [ A^{\alpha)} ,
A_{\alpha;\beta}{}^{\beta} ] ]
+ {1 \over 30} ( [ A^{(\rho} , [ A^{\alpha)} ,
A_{\beta;\alpha}{}^{\beta} ] ]
 \nonumber \\
\fl &&\quad
- {4 \over 135} [ A^{(\rho} , [ A^{\alpha)} ,
A^{\beta}{}_{;\beta\alpha} ] ]
-\, {1 \over 180} [ A_{(\alpha} , [ A_{\beta)} ,
A^{\alpha;\beta\rho} ] ]
+ {11 \over 180} [ A_{(\alpha} , [ A_{\beta)} ,
A^{\alpha;\rho\beta} ] ]
 \nonumber \\
\fl &&\quad 
-\, {1 \over 270} [ A_{(\alpha} , [ A_{\beta)} ,
A^{\rho;\alpha\beta} ] ]
+\, {1 \over 10} [ A^{\alpha}{}_{;\alpha} , [ A^{\rho} , 
A^{\beta}{}_{;\beta} ] ]
+ {23 \over 180} [ A^{\alpha}{}_{;\alpha} , [ A_{\beta} , 
A^{\beta;\rho} ] ] 
 \nonumber \\
\fl &&\quad 
- {11 \over 15} [ A^{\alpha;\beta} , [ A^{\rho} , 
A_{\alpha;\beta} ] ]
-\, {1 \over 30} [ A^{\alpha;\beta} , [ A^{\rho} , 
A_{\beta;\alpha} ] ]
+\, {1 \over 6} [ A_{\alpha;\beta} , [ A^{\alpha} , A^{\rho;\beta} ] ] 
\nonumber \\
\fl &&\quad 
+ {1 \over 60} [ A_{\alpha;\beta} , [ A^{\beta} , A^{\rho;\alpha} ] ] 
-\, {2 \over 45} [ A_{\alpha;\beta} , [ A^{\beta} , A^{\alpha;\rho} ] ]
-\, {14 \over 45} [ A_{\alpha;\beta} , [ A^{\alpha} , A^{\beta;\rho} ] ]
\nonumber \\
\fl &&\quad 
+\, {1 \over 30} [ A^{\rho;\alpha} , [ A_{\alpha} , 
A^{\beta}{}_{;\beta} ] ] 
+ {1 \over 15} [ A^{\rho;\alpha} , [ A^{\beta} , 
A_{\beta;\alpha} ] ] 
- {1 \over 15} [ A^{\rho;\alpha} , [ A^{\beta} , A_{\alpha;\beta} ] ]
\nonumber \\
\fl &&\quad 
+\, {1 \over 36} [ A^{\alpha;\rho} , [ A_{\alpha} , 
A^{\beta}{}_{;\beta} ] ] 
+ {1 \over 18} [ 
A^{\alpha;\rho} , [ A^{\beta} , A_{\alpha;\beta} ] ] 
+ {1 \over 36} [ A^{\alpha;\rho} , [ A^{\beta} , A_{\beta;\alpha} ] ] 
 \bigg)_{;\rho} \nonumber \\
\fl &&
+\, {1 \over 30} [ A_{\alpha} , \{ \{ A^{\alpha} , A^{\beta} \}  , 
F_{\beta\mu}{}^{;\mu} \} ] 
+\, {4 \over 15} [ A^{\alpha} , \{ A_{\beta} A^{\beta} , 
F_{\alpha\mu}{}^{;\mu} \} ]
\nonumber \\
\fl &&
+ {1 \over 2} [ A_{\alpha} , [ [ A^{\alpha} , A^{\beta} ] , 
F_{\beta\mu}{}^{;\mu}  ] ]
-\, {7 \over 90} [ A_{\alpha} , [ A^{\alpha} , [ A^{\beta}  ,
F_{\beta\mu}{}^{;\mu} ] ] ] 
-\, {5 \over 18} [ \{ A_{\alpha} A^{\alpha} , A^{\beta} \} , 
F_{\beta\mu}{}^{;\mu} ]
\nonumber \\ 
\fl &&
+\, {1 \over 9} [ A_{\alpha} A^{\beta} A^{\alpha} ,
F_{\beta\mu}{}^{;\mu} ]
+\, {17 \over 30} [ A_{\alpha} , [ A^{\alpha} , [ A^{\beta} , 
F_{\beta\mu} ]^{;\mu} ] ]
+\, {1 \over 10} [ A^{(\alpha} , [ A^{\beta)} , [ A_{\alpha} , 
F_{\beta\mu}{}^{;\mu} ] ] ]
\nonumber \\
\fl &&
- {1 \over 90} [ A^{(\alpha} , [ A^{\beta)} , [ A_{\alpha} , 
F_{\beta\mu} ]^{;\mu} ] ]
+\, {1 \over 20} [ A^{\alpha} , \{ \{ A^{\beta} , A^{\gamma} \} , 
F_{\alpha\beta;\gamma} \} ]
- {1 \over 2} [ A^{\alpha} , [ [ A^{\beta} , A^{\gamma} ] , 
F_{\alpha\beta;\gamma} ] ]
\nonumber \\
\fl &&
+\, {2 \over 9} [ [ A^{\alpha} A^{\beta} , A^{\gamma} ] , 
F_{\alpha\beta;\gamma} ]
-\, {1 \over 90} [ A^{(\alpha} , [ A^{\beta)} , [ A^{\rho} , 
F_{\rho\alpha} ]_{;\beta} ] ]
- {1 \over 540} [ A^{\alpha} , [ A_{\lambda}{}^{;\lambda} , 
[ A^{\beta}, F_{\alpha\beta} ] ] ]
\nonumber \\
\fl &&
+\, {1 \over 20} [ A^{\alpha} , \{ \{ A^{\beta} , 
A_{\lambda}{}^{;\lambda} \} , F_{\alpha\beta} \} ]
- {2 \over 5} [ A^{\alpha} A^{\beta} , [ A_{\lambda}{}^{;\lambda} ,
F_{\alpha\beta} ] ]
+\, {3 \over 5} [ A_{\alpha} , 
F^{\alpha\beta} A_{\beta} A_{\lambda}{}^{;\lambda} ] 
\nonumber \\
\fl &&
+\, {3 \over 5} [ A_{\alpha} , A_{\lambda}{}^{;\lambda} A_{\beta} F^{\alpha\beta} ] 
+ {1 \over 12} [ A^{\alpha} , [ [ A^{\beta} , A_{\lambda}{}^{;\lambda} ] , F_{\alpha\beta} ] ]
+ {67 \over 270} [  A_{\lambda}{}^{;\lambda} , [ A^{\alpha} A^{\beta} ,
F_{\alpha\beta} ] ]
\nonumber \\ 
\fl &&
+ {1 \over 15} [ A^{\lambda} , \{ \{ A_{\lambda} , A^{\alpha;\beta} \} , F_{\alpha\beta} \} ] 
+ {1 \over 5} [ A_{\lambda} A^{\lambda} , \{ A^{\alpha;\beta} , F_{\alpha\beta} \} ]
+ {1 \over 4} [ A_{\lambda} , [ [ A^{\lambda} , A^{\alpha;\beta} ] , 
F_{\alpha\beta} ] ]
\nonumber \\
\fl &&
+ {7 \over 15} [ A^{\alpha;\beta} , \{  A_{\lambda} A^{\lambda} , F_{\alpha\beta} \} ]
- {4 \over 15} [ A^{\lambda} , [ A_{\lambda} , [ A^{\alpha;\beta} ,
F_{\alpha\beta} ] ] ] 
+ {2 \over 45}  [ A^{(\alpha} , [ A^{\lambda)} , 
[ A^{\beta}{}_{;\lambda} , F_{\alpha\beta} ] ] ] \nonumber \\ 
\fl &&
+\, {1 \over 30} [ A^{\alpha;\lambda} , [ \{ A^{\beta} , A_{\lambda} \} , 
F_{\alpha\beta} ] ] 
+ {1 \over 30} [ A_{\lambda} A^{\alpha} ,
A^{\beta;\lambda} F_{\alpha\beta} ] 
+ {1 \over 30} [ A^{\alpha} A_{\lambda} , 
F_{\alpha\beta}  A^{\beta;\lambda} ]
\nonumber \\
\fl &&
+\, {1 \over 15} [ A^{[\lambda|} , 
F_{\alpha\beta} A^{|\alpha]} A^{\beta}{}_{;\lambda} 
+ A^{\beta}{}_{;\lambda} A^{|\alpha]} F_{\alpha\beta} ]
+ {1 \over 20} [ A^{\alpha} , \{ \{ A_{\lambda} , A^{\beta;\lambda} \} , F_{\alpha\beta}  \} ]
\nonumber \\
\fl &&
-\, {1 \over 4} [ A^{\alpha;\lambda} , [ 
[ A^{\beta} , A_{\lambda} ] , F_{\alpha\beta} ] ]
+ {1 \over 4} [ A^{\alpha} , [
[ A_{\lambda} , A^{\beta;\lambda} ] , F_{\alpha\beta} ] ]
+\, {1 \over 30} [ A^{\alpha;\lambda} ,
\{ \{ A^{\beta} , A_{\lambda} \} , F_{\alpha\beta} \} ]
\nonumber \\
\fl &&- {1 \over 20} [ \{ A^{\alpha} , A_{\lambda} \} ,
\{ A^{\beta;\lambda} , F_{\alpha\beta} \} ]
- {1 \over 540} ( [ A^{\alpha} , [ A^{\beta;\lambda} , [ A_{\lambda} , F_{\alpha\beta} ] ] ]
- [ A^{\alpha;\lambda} , [ A^{\beta} , [ A_{\lambda} , F_{\alpha\beta} ] ] ] )
\nonumber \\
\fl &&
+\, {13 \over 60} [ A_{\lambda} , \{ 
\{ A^{\alpha} , A^{\lambda;\beta} \} , F_{\alpha\beta} \} ]
-\, {1 \over 10} [ A^{\lambda;\alpha} , 
\{ \{ A^{\beta} , A_{\lambda} \} , F_{\alpha\beta} \} ]
+ {1 \over 15} [ A_{\lambda} A^{\alpha} , A^{\lambda;\beta} F_{\alpha\beta} ]
\nonumber \\
\fl &&
+ {1 \over 15} [ A^{\alpha} A_{\lambda} , F_{\alpha\beta} A^{\lambda;\beta} ]
+ {1 \over 30} [ A^{[\lambda|} , 
F_{\alpha\beta} A^{|\alpha]} A_{\lambda}{}^{;\beta}
+ A_{\lambda}{}^{;\beta} A^{|\alpha]} F_{\alpha\beta} ]
\nonumber \\
\fl &&
+\, {1 \over 20} [ A^{\alpha} , \{
\{ A_{\lambda} , A^{\lambda;\beta} \} , F_{\alpha\beta} \} ]
+\, {1 \over 30} [ A^{\alpha} A_{\lambda} , 
A^{\lambda;\beta} F_{\alpha\beta} ]
+\, {1 \over 30} [ A_{\lambda} A^{\alpha} , 
F_{\alpha\beta} A^{\lambda;\beta} ] 
\nonumber \\
\fl &&
+ {1 \over 2} [ A^{\lambda;\alpha} , [ [ A^{\beta} , A_{\lambda} ] , 
F_{\alpha\beta} ] ]
- {1 \over 4} [ A_{\lambda} , [ [ A^{\alpha} , A^{\lambda;\beta} ] , 
F_{\alpha\beta} ] ]
\nonumber \\
\fl &&
-\, {3 \over 20} [ \{ A^{\alpha} , A_{\lambda} \} , 
\{ A^{\lambda;\beta} , F_{\alpha\beta} \} ]
- {1 \over 15} [ A^{(\alpha} , [ A^{\lambda)} , 
[ A_{\lambda}{}^{;\beta} , F_{\alpha\beta} ] ] ]
\nonumber \\
\fl &&
+\, \Big[ {1 \over 4} \{ A_{\lambda} , A^{\lambda;\alpha} \} 
+  {1 \over 36} \{ A_{\lambda} , A^{\alpha;\lambda} \}  
-  {5 \over 18} \{ A^{\alpha} , A_{\lambda}{}^{;\lambda} \} , 
\{ A^{\beta} , F_{\alpha\beta} \} \Big]
\nonumber \\
\fl &&
+\, \Big[ {1 \over 36} [ A_{\lambda} , A^{\alpha;\lambda} ]  
- {1 \over 9} [ A_{\lambda} , A^{\lambda;\alpha} ] 
-  {1 \over 12} [ A^{\alpha} , A_{\lambda}{}^{;\lambda} ] , 
[ A^{\beta} , F_{\alpha\beta} ] \Big]
\nonumber \\
\fl &&
+\, \Big[ -\, {1 \over 3} \{ A^{\gamma} , A^{\alpha;\beta} \} 
+ {1 \over 36} \{ A^{\alpha} , A^{\beta;\gamma} \}  
+  {1 \over 12} \{ A^{\alpha} , A^{\gamma;\beta} \}  , 
\{ A_{\gamma} , F_{\alpha\beta} \} \Big]
\nonumber \\
\fl &&
+\, \Big[ {1 \over 6} [ A^{\gamma} , A^{\alpha;\beta} ] 
+ {1 \over 4} [ A^{\alpha} , A^{\beta;\gamma} ] 
- {1 \over 6} [ A^{\alpha} , A^{\gamma;\beta} ]  , 
[ A_{\gamma} , F_{\alpha\beta} ] \Big]
\nonumber \\
\fl &&
+\, R^{;\alpha} \Big(
- {3 \over 10} [ A_{\alpha} , A_{\beta} A^{\beta} ] 
+ {7 \over 60} [ A_{\beta} , A_{\alpha} A^{\beta} ] 
- {7 \over 72} [ A_{\beta} , [ A^{\beta} , A_{\alpha} ] ] \Big)
\nonumber \\
\fl &&
+\, R^{\alpha\beta;\gamma} \Big(
{4 \over 45} [ A_{\alpha} A_{\beta} , A_{\gamma} ] 
+ {13 \over 36} [ A_{\alpha} A_{\gamma} , A_{\beta} ]  
- {37 \over 180} [  A_{\gamma} A_{\alpha} , A_{\beta} ] \Big)
\nonumber \\
\fl &&
+\, R \Big( 
- {1 \over 36} [ A^{\alpha}{}_{;\alpha} , A_{\beta} A^{\beta} ] 
- {1 \over 9} [ A_{\beta} ,  A^{\alpha}{}_{;\alpha} A^{\beta} ] 
- {1 \over 36} [ A_{\beta} , [ A^{\beta} , A^{\alpha}{}_{;\alpha} ] ] 
\nonumber \\
\fl &&\qquad 
-\, {1 \over 9} [ A_{\alpha} , A_{\beta} A^{\alpha;\beta} ]
- {1 \over 9} [ A^{\alpha;\beta} , A_{\beta} A_{\alpha} ] 
+ {1 \over 18} [ A_{\alpha} A^{\alpha;\beta} , A_{\beta} ]  \Big)
\nonumber \\
\fl &&+\, R^{\alpha\beta} \Big(
{7 \over 90} [ A_{\alpha} A_{\beta} , A_{\gamma}{}^{;\gamma} ] 
+ {2 \over 15}  [ A_{\alpha} A_{\gamma}{}^{;\gamma} , A_{\beta} ] 
+ {7 \over 135} [ A_{\alpha} , [ A_{\beta} , A_{\gamma}{}^{;\gamma} ] ]
\nonumber \\
\fl &&\qquad
-\, {25 \over 36} [ A_{\alpha;\beta} , A_{\gamma} A^{\gamma} ]
+ {7 \over 45} [ A_{\gamma} , A_{\alpha;\beta} A^{\gamma} ] 
- {7 \over 90} [ A_{\gamma} , [ A^{\gamma} , A_{\alpha;\beta} ] ]  
\nonumber \\ 
\fl && \qquad
+\, {23 \over 180} [ [ A_{\alpha;\gamma} , A^{\gamma} ] , A_{\beta} ]
- {1 \over 45} [ A_{\gamma} , A_{\alpha} A_{\beta}{}^{;\gamma} ] 
+ {1 \over 36} [ A_{\alpha;\gamma} , [ A_{\beta} , A^{\gamma} ] ]  
\nonumber \\
\fl &&\qquad
+\, {7 \over 180} [ A^{\gamma} , A_{\gamma;\alpha} A_{\beta}  ]
+ {1 \over 2} [ A^{\gamma} , A_{\alpha} A_{\gamma;\beta}] 
+ {7 \over 12} [ A_{\gamma;\alpha} , A_{\beta} A^{\gamma} ]  
\nonumber \\
\fl &&\qquad 
-\, {1 \over 90} [ A_{\gamma;\alpha} , A^{\gamma} A_{\beta} ] 
+ {1 \over 5} [ A_{\gamma;\alpha} A^{\gamma} , A_{\beta} ]  
- {11 \over 180} [ A^{\gamma} A_{\alpha;\gamma} , A_{\beta} ] \Big)
\nonumber \\
\fl && 
+\, R^{\alpha\beta\gamma\delta} \Big( 
-\, {1 \over 270} [ A_{\alpha} A_{\beta} , A_{\gamma;\delta} ] 
+\, {3 \over 10} [ A_{\gamma} , A_{\beta;\delta} A_{\alpha} ]  
-\, {23 \over 180} [ A_{\gamma} , A_{\alpha} A_{\beta;\delta} ]
\nonumber \\
\fl &&\qquad
-\, {5 \over 36} [ A_{\alpha} A_{\gamma} , A_{\beta;\delta} ] 
- {11 \over 135} [ A_{\beta;\delta} A_{\gamma} , A_{\alpha} ]  
- {31 \over 540} [ A_{\gamma} A_{\beta;\delta} , A_{\alpha} ] \Big)
\nonumber \\
\fl && 
+{\rm i}\, \epsilon^{\alpha\beta\gamma\delta\epsilon\zeta} \Big( 
{1 \over 10} [ A_{\alpha} A_{\beta;\gamma} , A_{\delta;\epsilon} A_{\zeta} ] 
+\, {1 \over 12} [ A_{\alpha;\beta} , A_{\gamma;\delta} A_{\epsilon} A_{\zeta} ]
-\, {1 \over 15} [ A_{\alpha;\beta} A_{\gamma;\delta} , A_{\epsilon} A_{\zeta} ] 
\nonumber \\ 
\fl && \qquad
-\, {1 \over 30} [ A_{\alpha} , A_{\beta;\gamma} A_{\delta} A_{\epsilon;\zeta} ] 
-\, {1 \over 40} [ A_{\alpha} A_{\beta;\gamma} , A_{\delta} A_{\epsilon;\zeta} ] 
+\, {4 \over 15} [ A_{\alpha} A_{\beta} A_{\gamma;\epsilon\zeta} , A_{\delta} ]
\nonumber \\
\fl && \qquad
-\, {4 \over 15} [ A_{\alpha} A_{\beta;\epsilon\zeta} , A_{\gamma} A_{\delta} ] 
-\, {2 \over 15} [ A_{\alpha;\epsilon\zeta} , A_{\beta} A_{\gamma} A_{\delta} ] 
+\, {19 \over 40} [ A_{\alpha} A_{\beta} A_{\gamma} F_{\epsilon\zeta} , A_{\delta} ]
\nonumber \\
\fl && \qquad
+\, {29 \over 60} [ A_{\alpha} A_{\beta} F_{\epsilon\zeta} , A_{\gamma} A_{\delta} ] 
+\, {11 \over 30} [ A_{\alpha} F_{\epsilon\zeta} , A_{\beta} A_{\gamma} A_{\delta} ] 
+\, {7 \over 30} [ F_{\epsilon\zeta} , A_{\alpha} A_{\beta} A_{\gamma} A_{\delta} ] \Big)
\nonumber \\
\fl && 
-\, {31 \over 60} [ A^{\lambda} {}_{;\lambda} , A_{\alpha}  A^{\alpha} A_{\beta}  A^{\beta} ] 
-\, {19 \over 16} [ A_{\alpha}  A^{\lambda} {}_{;\lambda} , A_{\beta} A^{\beta} A^{\alpha} ] 
+\, {26 \over 15} [ A_{\alpha} A^{\alpha} A^{\lambda} {}_{;\lambda} ,
A_{\beta} A^{\beta}  ] 
\nonumber \\ 
\fl && 
-\, {1 \over 4} [ A_{\alpha} A_{\beta} A^{\beta} A^{\lambda} {}_{;\lambda} , A^{\alpha} ] 
+\, {5 \over 18} [ A^{\lambda} {}_{;\lambda} A_{\alpha} , A^{\alpha} A_{\beta}  A^{\beta} ] 
-\, {13 \over 12} [ A_{\alpha}{}_{;\lambda} A^{\lambda} , A_{\beta}  A^{\beta} A^{\alpha} ] 
\nonumber \\ 
\fl && 
+\, {58 \over 15} [ A_{\alpha} A^{\alpha;\lambda} A_{\lambda} , A_{\beta}  A^{\beta} ] 
-\, {97 \over 60} [ A^{\alpha} A^{\beta}  A_{\beta;\lambda} A^{\lambda} , A_{\alpha} ]  
-\, {47 \over 30} [ A^{\lambda} , A_{\alpha}  A^{\alpha} A^{\beta} A_{\beta;\lambda} ] 
\nonumber \\ 
\fl && 
-\, {7 \over6} [ A_{\alpha} A^{\alpha;\lambda} , A_{\lambda} A_{\beta}  A^{\beta} ] 
+\, {53 \over 15} [ A_{\alpha;\lambda} A^{\alpha} A^{\lambda} , A_{\beta}  A^{\beta} ] 
-\, {8 \over 15} [ A^{\alpha} A_{\beta;\lambda} A^{\beta} A^{\lambda} , A_{\alpha} ] 
\nonumber  \\
\fl && 
-\, {21 \over 10} [ A^{\lambda} , A_{\alpha} A^{\alpha} A_{\beta;\lambda} A^{\beta} ] 
-\, {31 \over 30} [ A^{\alpha} A^{\lambda} , A_{\beta} A^{\beta} A_{\alpha;\lambda} ] 
-\, {11 \over 6} [ A_{\alpha;\lambda} A^{\alpha} , A^{\lambda} A_{\beta} A^{\beta} ] 
\nonumber \\ 
\fl && 
-\, {5 \over 12} [ A^{\alpha} A_{\beta;\lambda} , A^{\beta} A^{\lambda} A_{\alpha} ] 
-\, {31 \over 30} [ A_{\alpha;\lambda} A_{\beta} A^{\beta} A^{\lambda} , A^{\alpha} ] 
+\, {1 \over 3} [ A^{\lambda} , A^{\alpha} A_{\alpha;\lambda} A_{\beta} A^{\beta} ] 
\nonumber \\ 
\fl && 
-\, {8 \over 15} [ A^{\alpha} A^{\lambda} , A^{\beta} A_{\beta;\lambda} A_{\alpha} ] 
+\, {17 \over 10} [ A_{\alpha} A^{\alpha} A^{\lambda} , A^{\beta} A_{\beta;\lambda} ]
+\, {5 \over 12} [ A_{\alpha;\lambda} A_{\beta} , A^{\beta} A^{\lambda} A^{\alpha} ]  
\nonumber \\
\fl &&
+\, {101 \over 60} [ A^{\lambda} , A_{\alpha;\lambda} A^{\alpha} A_{\beta} A^{\beta} ] 
-\, {97 \over 60} [ A_{\alpha} A^{\lambda} , A_{\beta;\lambda} A^{\beta} A^{\alpha} ] 
+\, {27 \over 10} [ A_{\alpha} A^{\alpha} A^{\lambda} , A_{\beta;\lambda} A^{\beta} ] 
\nonumber \\ 
\fl && 
-\, {13 \over 6} [ A^{\beta} A_{\alpha} A^{\alpha} A^{\lambda} , A_{\beta;\lambda} ] 
-\, {13 \over 12} [ A^{\lambda} A_{\alpha;\lambda} , A^{\alpha} A_{\beta} A^{\beta} ] 
-\, {1 \over 4} [ A^{\lambda} {}_{;\lambda} , A_{\alpha}  A_{\beta}  A^{\alpha} A^{\beta} ] 
\nonumber \\
\fl && 
+\, {62 \over 45} [ A_{\alpha} A^{\lambda} {}_{;\lambda} , A_{\beta} A^{\alpha} A^{\beta} ] 
-\, {5 \over 2} [ A_{\alpha} A_{\beta} A^{\lambda} {}_{;\lambda} , A^{\alpha} A^{\beta} ] 
+\, {8 \over 3} [ A_{\alpha} A_{\beta} A^{\alpha} A^{\lambda} {}_{;\lambda} , A^{\beta} ] 
\nonumber \\ 
\fl && 
+\, {2 \over 9} [ A^{\lambda} {}_{;\lambda} A_{\alpha} , A_{\beta}  A^{\alpha} A^{\beta} ] 
+\, {101 \over 90} [ A_{\alpha;\lambda} A^{\lambda} , A_{\beta}  A^{\alpha}  A^{\beta} ] 
-\, {43 \over 15} [ A_{\alpha} A_{\beta;\lambda} A^{\lambda} , A^{\alpha} A^{\beta} ] 
\nonumber \\ 
\fl && 
+\, {7 \over 4} [ A_{\alpha} A_{\beta} A^{\alpha;\lambda} A_{\lambda} , A^{\beta} ] 
+\, {7 \over 20} [ A_{\lambda} , A_{\alpha} A_{\beta} A^{\alpha} A^{\beta;\lambda} ] 
+\, {2 \over 3} [ A_{\alpha} A_{\beta;\lambda} , A^{\lambda} A^{\alpha} A^{\beta} ] 
\nonumber \\ 
\fl && 
-\, {12 \over 5} [ A_{\alpha} {}_{;\lambda} A_{\beta} A^{\lambda} , A^{\alpha} A^{\beta} ] 
+\, {11 \over 10} [ A_{\alpha} A_{\beta;\lambda} A^{\alpha} A^{\lambda} , A^{\beta} ] 
+\, {9 \over 20} [ A_{\lambda} , A_{\alpha} A_{\beta} A^{\alpha;\lambda} A^{\beta} ] 
\nonumber \\ 
\fl &&
+\, {4 \over 3} [ A_{\alpha} A_{\lambda} , A_{\beta} A^{\alpha} A^{\beta;\lambda} ] 
+\, {11 \over 12} [ A_{\alpha} {}_{;\lambda} A_{\beta} , A^{\lambda} A^{\alpha} A^{\beta} ] 
+\, {5 \over 12} [ A_{\alpha} A_{\beta;\lambda} , A^{\alpha} A^{\lambda} A^{\beta} ] 
\nonumber \\ 
\fl &&
+\, {4 \over 5} [ A_{\alpha;\lambda} A_{\beta} A^{\alpha} A^{\lambda} , A^{\beta} ] 
-\, {49 \over 60} [A_{\lambda} , A_{\alpha} A_{\beta} {}^{;\lambda} A^{\alpha} A^{\beta} ] 
-\, {2 \over 5} [ A_{\alpha} A_{\lambda} , A_{\beta} A^{\alpha;\lambda} A^{\beta} ] 
\nonumber \\ 
\fl &&
-\, {89 \over 60} [  A_{\alpha} A_{\beta} A_{\lambda} , A^{\alpha}  A^{\beta;\lambda} ] 
-\, {5 \over 12} [ A_{\alpha;\lambda} A_{\beta} , A^{\alpha} A^{\lambda} A^{\beta} ] 
-\, {61 \over 30} [ A^{\lambda} , A_{\alpha;\lambda} A_{\beta} A^{\alpha} A^{\beta} ] 
\nonumber \\ 
\fl &&
+\, {25 \over 12} [ A_{\alpha} A^{\lambda} , A_{\beta;\lambda} A^{\alpha} A^{\beta} ] 
-\, {151 \over 60} [ A_{\alpha} A_{\beta} A_{\lambda} , A^{\alpha;\lambda} A^{\beta} ] 
+\, {49 \over 20} [ A_{\alpha} A^{\beta} A^{\alpha} A^{\lambda} , A_{\beta;\lambda} ] 
\nonumber \\ 
\fl &&
+\,  [ A^{\lambda} A_{\alpha;\lambda} , A_{\beta} A^{\alpha} A^{\beta} ] 
+\, {11 \over 15} [ A^{\lambda} {}_{;\lambda} , A_{\alpha}  A_{\beta} A^{\beta}  A^{\alpha} ] 
-\, {149 \over 180} [ A^{\alpha} A^{\lambda} {}_{;\lambda} , A_{\alpha}  A_{\beta} A^{\beta} ] 
\nonumber \\ 
\fl && 
+\, {61 \over 30} [ A_{\alpha}  A_{\beta} A^{\lambda} {}_{;\lambda} , A^{\beta}  A^{\alpha} ] 
-\, {31 \over 20} [ A_{\alpha} A^{\alpha} A_{\beta} A^{\lambda} {}_{;\lambda} , A^{\beta} ] 
-\, {13 \over 18} [ A^{\lambda} {}_{;\lambda} A_{\alpha} , A_{\beta} A^{\beta}  A^{\alpha} ] 
\nonumber \\ 
\fl && 
-\, {43 \over 30} [ A^{\alpha;\lambda} A_{\lambda} , A_{\alpha}  A_{\beta}  A^{\beta} ] 
+\, {35 \over 12} [ A_{\alpha}  A_{\beta} {}_{;\lambda} A^{\lambda} , A^{\beta}  A^{\alpha} ] 
-\, {31 \over 15} [ A_{\alpha}  A^{\alpha} A_{\beta;\lambda} A^{\lambda} , A^{\beta} ] 
\nonumber \\
\fl && 
-\, {7 \over 60} [ A_{\lambda} , A_{\alpha}  A_{\beta}  A^{\beta} A^{\alpha;\lambda} ] 
-\, {5 \over 12} [ A_{\alpha} A_{\beta} {}_{;\lambda} , A^{\lambda} A^{\beta}  A^{\alpha} ] 
+\, {32 \over 15} [ A_{\alpha;\lambda} A_{\beta} A^{\lambda} , A^{\beta} A^{\alpha} ] 
\nonumber \\
\fl && 
-\, {5 \over 6} [ A_{\alpha} A^{\alpha;\lambda}  A_{\beta} A_{\lambda} , A^{\beta} ] 
-\, {13 \over 60} [ A_{\lambda} , A_{\alpha} A_{\beta} A^{\beta;\lambda} A^{\alpha} ] 
\nonumber \\ 
\fl && 
-\, {47 \over 30} [ A_{\alpha} A^{\lambda} , A^{\alpha} A^{\beta} A_{\beta;\lambda} ] 
-\, {2 \over 3} [ A_{\alpha;\lambda} A_{\beta} , A^{\lambda} A^{\beta} A^{\alpha} ] 
-\, {2 \over 3} [ A_{\alpha} A^{\alpha;\lambda} , A_{\beta} A_{\lambda} A^{\beta} ] 
\nonumber \\
\fl && 
-\, {47 \over 30} [ A_{\alpha;\lambda} A^{\alpha} A^{\beta} A^{\lambda} , A_{\beta} ] 
+\, {53 \over 60} [ A^{\lambda} A^{\alpha} A^{\beta} {}_{;\lambda} A_{\beta} A_{\alpha} ] 
-\, {5 \over 6} [ A_{\alpha} A_{\lambda} , A^{\alpha} A^{\beta;\lambda} A_{\beta} ] 
\nonumber \\ 
\fl && 
+\, {22 \over 15} [ A_{\alpha} A_{\beta} A_{\lambda} , A^{\beta} A^{\alpha;\lambda} ] 
+\, {2 \over 3} [ A_{\alpha;\lambda} A^{\alpha} , A^{\beta} A^{\lambda} A_{\beta} ] 
+\, {47 \over 20} [ A^{\lambda} A_{\alpha;\lambda} A_{\beta} A^{\beta} , A^{\alpha} ] 
\nonumber \\ 
\fl && 
-\, {31 \over 15} [ A_{\alpha} A_{\lambda} , A^{\alpha;\lambda} A_{\beta} A^{\beta} ] 
+\, {5 \over 2} [ A_{\alpha} A_{\beta} A_{\lambda} , A^{\beta;\lambda} A^{\alpha} ] 
-\, {161 \over 60} [ A_{\alpha} A^{\alpha} A^{\beta} A^{\lambda} , A_{\beta;\lambda} ] 
\nonumber \\ 
\fl && 
-\, {5 \over 4} [ A^{\lambda} A_{\alpha;\lambda} , A_{\beta} A^{\beta} A^{\alpha} ] 
+\,  {{\rm i} \over 30} \, \epsilon^{\alpha\beta\gamma\delta\kappa\lambda}
A_{\alpha}  A_{\beta} A_{\gamma} A_{\delta} A_{\kappa} A_{\lambda} \bigg]. \label{eq:A6com}
\end{eqnarray}
Many commutator terms may be rewritten 
by using the Jacobi identity of the commutator
and the commutation relation of the covariant differentiation. 
These changes do not vary the degree of $A_{\mu}$. 
The last term with $A_{\mu}$ of six degrees forms
 a commutator factor $\epsilon^{\alpha\beta\gamma\delta\kappa\lambda}
\,{\rm tr}\, T^a [ A_{\alpha} , A_{\beta} A_{\gamma} A_{\delta} A_{\kappa} A_{\lambda} ]$.

\section{Discussion}
In the model in which the fermion interacts with the vector and the
axial-vector fields in curved space, 
imposing the conservation equation of the vector current
at the quantum level,
the gauge anomalies in four and six dimensions  
are represented in tensorial form. 
The gauge anomaly 
in the model in four-dimensional flat space 
has already been given~\cite{Bardeen,GA1,GA2,GA3a,GA3b,GA4}. 
In curved space, the expression (\ref{eq:famous}) 
of $G^{(4)a}$ contains
terms described by contraction of $R^{ab}{}_{\mu\nu}$ 
and a covariant derivative of $A_{\mu}$,
and is rewritten as in (\ref{eq:GA4}).
If only $T^a$ written visibly in (\ref{eq:GA}) is a unit matrix, then
the gauge anomaly agrees with the chiral U(1) anomaly. 
When all $T^a$ do not commute each other, 
the gauge anomaly should have 
the chiral U(1) anomaly-like part, 
which is expressed by the contraction of 
the curvature 2-form $R^{ab}$, 
the trace of the strength 2-form $F$ of the vector field $V_{\mu}$
and the total derivative terms containing $A_{\mu}$, 
due to the index theorem~\cite{IndexTheorem1,IndexTheorem2}.
The other part of the gauge anomaly becomes terms 
containing the commutator of $A_{\mu}$, $F_{\alpha\beta}$ 
and their derivatives, 
because the commutators should disappear
in the absence of the above $T^a$.
The non-derivative term containing $A_{\mu}$ of the same degree as
the spacetime dimension $2n$ is expressed by a commutator factor 
$\epsilon^{\mu_1\mu_2\cdots \mu_{2n}}
{\rm tr}\, T^a [ A_{\mu_1} , A_{\mu_2} \cdots A_{\mu_{2n}} ]$,
because of the contraction of the Levi-Civita symbol 
with the product of $A_{\mu}$.  
Indeed, $G^{(6)a}$ consists of 
the leading terms containing $R^{ab}{}_{\mu\nu}$ and $F_{\mu\nu}$,
the total derivative part in (\ref{eq:A6}) 
and the commutator part $G^{a}_{\rm com}$ in $(\ref{eq:A6com})$.  

The Lagrangian in $(\ref{eq:action})$ can be rewritten 
by using right- and left-handed Weyl fermions:
\begin{eqnarray}
&& {\cal L} = h\, \left(
\bar{\psi}_R\, {\rm i} {\ooalign{\hfil/\hfil\crcr${D}$}}_R \psi_R
+ \bar{\psi}_L\, {\rm i} {\ooalign{\hfil/\hfil\crcr${D}$}}_L \psi_L 
\right),\qquad 
P_{\pm} = {1 \pm \gamma_{2n+1} \over 2}, 
\nonumber \\
&&\psi_R = P_+ \psi,\quad 
{\ooalign{\hfil/\hfil\crcr${D}$}}_R = \gamma^{\mu}
(\nabla_{\mu} + R_{\mu})P_+ ,
\quad R_{\mu} \equiv V_{\mu} + A_{\mu}, \nonumber \\
&&\psi_L = P_- \psi, \quad 
{\ooalign{\hfil/\hfil\crcr${D}$}}_L = \gamma^{\mu}
(\nabla_{\mu} + L_{\mu})P_- ,
\quad L_{\mu} \equiv V_{\mu} - A_{\mu}. 
\end{eqnarray}
Though the Dirac operators ${\ooalign{\hfil/\hfil\crcr${D}$}}_R$ and
${\ooalign{\hfil/\hfil\crcr${D}$}}_L$ are not Hermitian, 
$\displaystyle{
{\ooalign{\hfil/\hfil\crcr${D}$}}_L^{\dag} 
= {\ooalign{\hfil/\hfil\crcr${D}$}}_R}$ and 
$\displaystyle{{\ooalign{\hfil/\hfil\crcr${D}$}}_R^{\dag} 
= {\ooalign{\hfil/\hfil\crcr${D}$}}_L}$. 
Then the Jacobian for ${\cal D}\bar{\psi}_L {\cal D}\psi_L$ 
in the path integral under the gauge transformation 
gives a half contribution of the gauge anomaly, 
which arises from the difference between the heat kernels regulated by 
${\ooalign{\hfil/\hfil\crcr${D}$}}_L {\ooalign{\hfil/\hfil\crcr${D}$}}_L^{\dag}$ 
and ${\ooalign{\hfil/\hfil\crcr${D}$}}_L^{\dag} {\ooalign{\hfil/\hfil\crcr${D}$}}_L$,
instead of ${\ooalign{\hfil/\hfil\crcr${D}$}}^2$ in $(\ref{eq:EA_GC})$.  
The Jacobian for ${\cal D}\bar{\psi}_R {\cal D}\psi_R$ yields the opposite sign of the contribution.
Though $V_{\mu}$ and $A_{\mu}$ of the anomaly 
in $(\ref{eq:famous})$-$(\ref{eq:A6com})$ are replaced 
by $R_{\mu}$ and $L_{\mu}$, respectively,
the form of $(R+L)_{\mu}$ in the anomaly is restricted, 
since $V_{\mu}$ appears 
only in the field strength $F_{\mu\nu}$ and the
covariant derivatives. However, the form of $(R-L)_{\mu}$ is no restriction, 
because the gauge symmetry for $A_{\mu}$ breaks down by the Wick rotation of the axial-gauge field,
by which ${\ooalign{\hfil/\hfil\crcr${D}$}}$ becomes Hermitian. 
The breaking of the symmetry spoils the gauge transformation of $R_{\mu}$ and $L_{\mu}$.

If all $A_{\mu}$ are Abelian in (\ref{eq:GA4}), then
$G^{a(4)}$ corresponds to the gauge anomaly in space with torsion, 
which is originally expressed by the third-order antisymmetric tensor.
The dual vector of the tensor in four dimensions behaves as the axial-vector~\cite{Torsion}. 
Then, there is no commutator term in (\ref{eq:GA4}).
Note that the dual tensor of torsion in six or higher dimensions 
is the third- or higher-order antisymmetric tensor.  
Therefore, the anomaly with the vector and the axial-vector fields 
in six-dimensional space with torsion 
will have new terms containing the third-order torsion tensor.

In supergravity~\cite{SG1,SG2}, 
the gravitino is described by 
the Rarita-Schwinger field $\psi_{\mu}$ 
with a suitably fixed gauge~\cite{GF1,GF2,GF3,GF4,GF5}, 
and some quantum effects for the gravitino are evaluated by  
the heat kernel for a fermion of spin $\frac{3}{2}$.
By treating the vector index `$\mu$' of $\psi_{\mu}$ 
as the internal index of a representation matrix for the Lie algebra of the special orthogonal group, 
the heat kernel for the gravitino can be applied like that for a fermion of spin $\frac{1}{2}$. 
When the chiral U(1) anomaly for the gravitino is calculated,
the strength $(T^a F^a_{\mu\nu})_{\alpha\beta}$ of the vector gauge field 
may be replaced by the curvature tensor $R_{\mu\nu\alpha\beta}$. 
However, since $T^a$ 
cannot exist by itself, there is no gauge anomaly 
for the gravitino of spin $\frac{3}{2}$.
In supergravity coupled with super Yang-Mills theory, the gauge anomaly for 
the gaugino of spin $\frac{1}{2}$ in curved space may be expressed by
the vector and the axial-vector bosons, 
which take the bilinear form of the gravitino and the gaugino, 
in place of $V_{\mu}$ and $A_{\mu}$, respectively.

\end{document}